\documentclass{iaus}
\usepackage{graphicx}
\usepackage{natbib}
\def\apj{{\it ApJ}}%
\def\aap{{\it A\&A}}%
\def\mnras{{\it MNRAS}}%
\def\pasp{{\it PASP}}%

\title[Stellar mass \& SED maps of galaxies] 
{Resolved maps of stellar mass and SED of galaxies from optical/NIR
imaging and SPS models}

\author[Zibetti, Charlot \& Rix]   
{Stefano Zibetti$^1$, St\'ephane Charlot$^2$ \and Hans-Walter Rix$^1$
}

\affiliation{$^1$Max-Planck-Institut f\"ur Astronomie, \\ K\"onigstuhl 17, 69117 Heidelberg, Germany \\ email: {\tt zibetti@mpia.de}, {\tt rix@mpia.de} \\[\affilskip]
  $^2$Institut d'Astrophysique de Paris, CNRS, Universit\'e Pierre \&
  Marie Curie,\\ 98 bis Boulevard Arago, 75014 Paris, France \\email:
  {\tt charlot@iap.fr}}

\pubyear{2009}
\volume{262}  
\pagerange{1--4}
\jname{Stellar Populations: Planning for the Next Decade}
\editors{S. Charlot, G. Bruzual, eds.}
\begin{document}

\maketitle

\begin{abstract}
  We report on the method developed by Zibetti, Charlot \& Rix (2009)
  to construct resolved stellar mass maps of galaxies from optical and
  NIR imaging. Accurate pixel-by-pixel colour information
  (specifically $g-i$ and $i-H$) is converted into stellar
  mass-to-light ratios with typical accuracy of 30\%, based on median
  likelihoods derived from a Monte Carlo library of 50,000 stellar
  population synthesis models that include dust and updated TP-AGB
  phase prescriptions. Hence, surface mass densities are computed.  In
  a pilot study, we analyze 9 galaxies spanning a broad range of
  morphologies. Among the main results, we find that: i) galaxies
  appear much smoother in stellar mass maps than at any optical or NIR
  wavelength; ii) total stellar mass estimates based on unresolved
  photometry are biased low with respect to the integral of resolved
  stellar mass maps, by up to 40\%, due to dust obscured regions being
  under-represented in global colours; iii) within a galaxy, {\it on
    local scales} colours correlate with surface stellar mass density;
  iv) the slope and tightness of this correlation reflect/depend on
  the morphology of the galaxy.

\end{abstract}

\firstsection 
\section{Introduction}
The key role of stellar mass to determine (or predict) the physical
properties of present day galaxies has been established by a number of
works since the last decade
\citep[e.g.][]{gavazzi+96,scodeggio+02,kauffmann+03b}. Indication has
been provided that stellar mass density might be even more fundamental
\citep{bell_dejong00,kauffmann+03b}. All these works, however, deal
with {\it global} estimates of the stellar mass-to-light ratio, which
is assumed to be uniform throughout a galaxy, contrary to what we
should expect based on well known spatial variations of stellar
population and dust properties. Hence, resolving the distribution of
stellar mass is crucial to properly measure total stellar mass and
mass density and to start investigating the structure and dynamics of
galaxies in an unbiased way. Having access to the mass distribution
also allows to address questions like: Is there any relation between
{\it local} stellar mass density and the {\it local} SED and physical
properties {\it within} a galaxy, similarly to the relations observed
{\it globally}? If so, what does it tell us about the internal
mechanisms of galaxy evolution?

With these goals and questions in mind, we have developed a new method
to build stellar mass maps of galaxies \citep[][ZCR09
hereafter]{ZCR09}, which we review in this contribution (Sec. 2). We
also present preliminary results on the colour-stellar mass density
relation within galaxies for a small sample of different morphological
types (Sec. 3).

\section{From multi-band imaging to stellar mass}
As described in ZCR09, our goal is to provide a
computationally fast and observationally cheap method to obtain
stellar mass maps.  We express the surface stellar mass density at any
given position $(\alpha,\delta)$ of a galaxy as:
\begin{equation}
  \Sigma_{M_*}(\alpha,\delta)=\Sigma_{\lambda}(\alpha,\delta) \Upsilon_{\lambda}(\alpha,\delta)
\end{equation}
where $\Sigma_{\lambda}(\alpha,\delta)$ is the surface brightness in
the chosen `luminance' band of effective wavelength $\lambda$ and
$\Upsilon_{\lambda}(\alpha,\delta)$ is the corresponding effective M/L
ratio. In turn, we want to express $\Upsilon_{\lambda}$ as a function
of colour indexes. Despite the diversity of stellar population
parameters and dust properties, if the luminance band and the colour
indexes are appropriately chosen, the scatter of $\Upsilon_{\lambda}$
as a function of colour(s) can be reasonably small
\citep[e.g. ][]{bell_dejong01}.  We choose the $H$ band (largely
equivalent to $J$ or $K$) as luminance band and we use $g-i$ and $i-H$
colour indexes to predict M/L$_H$. This choice is supported by the
fact that M/L variations in NIR are minimal with respect to shorter
wavelengths, while these two colours are the best combination in terms
of sensitivity and wavelength leverage. Based upon the 2007 version of
the \cite{BC03} code, which implements updated prescriptions for the
TP-AGB stellar evolutionary phase according to
\cite{marigo_girardi08}, we build a Monte Carlo library of 50,000
stellar population synthesis (SPS) models with a variety of star
formation histories, both continuous and bursty, metallicities and
dust attenuations \'a la \cite{charlot_fall00}. In this way we mean to
cover as uniformly as possible the parameter space occupied by
individual regions in galaxies on scales of $\approx 100$~pc, from
old, metal-rich bulges/spheroids, to young spiral arms with different
degrees of dust absorption. Then we bin the models in cells of
$0.05\times 0.05$~mag$^2$ in $g-i$ and $i-H$ and for each cell we
consider the median M/L$_H$. This look-up table is then used to assign
the median-likelihood M/L to each pixel in a galaxy, depending on its
colours. Our approach is therefore bayesian rather than frequentist
and, as such, depends to some extent on the chosen prior distribution
of models: a different choice of SFHs in particular can affect the M/L
of the bluest stellar populations (see ZCR09 for a detailed
discussion). We confirm \citep[e.g.][]{maraston05} that the treatment
of the TP-AGB phase in the SPS models has a great systematic impact on
the resulting colours and M/L in NIR, but this is mainly limited to
young/blue stellar populations.  Apart from systematic effects, we
find that the confidence half-range for the M/L is $\approx 30$\% over
most of the colour space, with the notable exception of the stellar
populations with ages $\lesssim 1$~Gyr, where the stellar evolution is
very rapid and results in larger ranges, up to a factor 2. The use of
two colours is crucial to obtain such accurate estimates: by
neglecting the dependence on $i-H$ one would easily
over/under-estimate the M/L by a factor 2 to 3.\footnote{In ZCR09 we
  show that using $i$-band and the $g-i$ colour as M/L predictor gives
  results that are in very good agreement with those based on $g-i-H$
  in most cases, except in stellar populations dominated by a very
  young burst or with very large extinction.}

To test our stellar mass mapping method we select a sample of 9 nearby
galaxies, which are also part of the SINGS survey
\citep{SINGS_kennicutt+03} and for which optical SDSS imaging and
medium-deep NIR imaging (either from GOLD Mine, \citealt{goldmine}, or
UKIDSS, \citealt{lawrence+07}) are available. The sample spans all
range of morphologies, from ellipticals to late spirals. As fluxes and
colours in individual pixels have to be accurate at few percent level
in order for our method to work, it is necessary to pre-process the
images to ensure that sufficiently high S/N can be reached also in the
regions with surface brightness as low as 24-25 mag$_g$
arcsec$^{-2}$. This is done using a newly developed adaptive smoothing
code {\sc adaptsmooth} (Zibetti in prep., ZCR09), which preserves the
maximum spatial resolution compatible with the requested S/N.

\begin{figure}[h]
\begin{center}
 \includegraphics[width=0.7\textwidth]{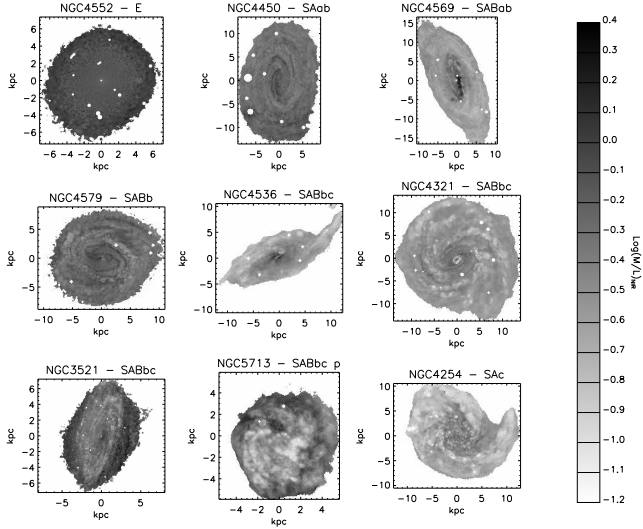} 
 \hspace*{6.0 cm}
 \caption{The maps of effective stellar mass to NIR light ratio of the
   9 galaxies, sorted by morphological type, from Elliptical to
   Sc. Darker grey represents higher M/L, as given by the scale
   aside.}
   \label{fig1}
\end{center}
\end{figure}

Figure \ref{fig1} shows the maps of effective $M_\ast/L_{NIR}$ for the
9 galaxies of the sample. Earlier type galaxies have more uniform and
higher M/L, on average. The young stellar populations that
characterize spiral arms result in lower M/L ratios, which in turn
decrease the spiral arm contrast. The resulting stellar mass maps
reveal that the structure of spiral galaxies is significantly smoother
than it appears at any optical or NIR wavelength. 
The colour gradients, especially visible in spiral galaxies, result in
M/L gradients, which in turn affect the structural parameters and make
them differ from the ones derived from light distribution. In
particular, the effective radius of spirals is smaller when the
stellar mass distribution is used.

In Fig. \ref{fig1} one can see lanes of much higher M/L (especially
NGC4569 and NGC4536), which correspond to dust lanes. These regions
can hide a significant amount of stellar mass, although the emerging
light hardly affects the global luminosity and colours of the
galaxy. As we show in ZCR09, this causes stellar masses obtained from
global photometry to be biased low with respect to what one gets from
integrating resolved stellar mass maps: the latter may miss up to 40\%
of the total stellar mass of a galaxy if dust obscured regions are
very extended (as in NGC4569 and NGC4536).

\section{Trends of colours with surface stellar mass density}
\begin{figure}[t]
\begin{center}
 \includegraphics[width=0.5\textwidth]{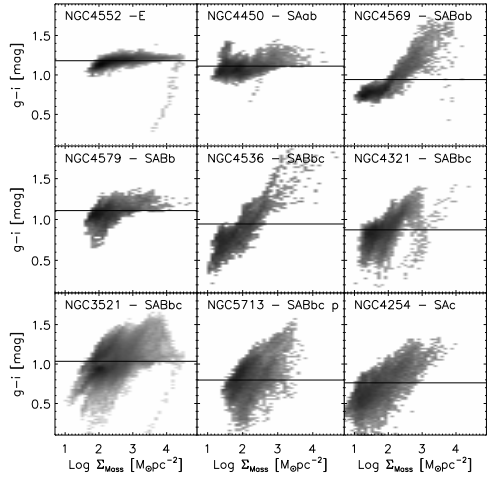} 
 \caption{The distribution of pixels in the $g-i$ vs surface stellar
   mass density plane for the 9 galaxies: in grey scale the log of
   normalized number of pixels per cell. The horizontal line marks the
   global colour of the galaxy.}
   \label{fig2}
\end{center}
\end{figure}

Based on the analysis of the previous section, we study the
correlation between colours (namely the optical $g-i$) and the surface
stellar mass density {\it within} each galaxy.  The distribution of
pixels as a function of these two quantities are shown in
Fig. \ref{fig2}, where the grey scale denotes the density of
pixels. As a general result, we see that the colour positively
correlates with surface stellar mass density: galaxies are redder in
higher density regions. Although this holds for all galaxies, the
slope and the dispersion of the correlation varies a lot along the
morphological sequence. For a typical Elliptical galaxy the relation
is very flat and very tight (consistent with a scatter due to
photometric errors only). As we move to later types the colour-mass
relation steepens, with low-density regions becoming increasingly
bluer, while the highest density regions have roughly constant red
colours. We can interpret this as a sign that star formation prefers
low density and that the colour-morphology relation is mainly set by
the relative weight of the younger stellar populations in the lower
density regions with respect to the red high-density `cores'. In
addition, we see that the scatter around the mean relation increases
going to later types, which have a more disomogeneous distribution of
physical properties at given surface mass density. The presence of
dust also increases the scatter, especially at high surface mass
density. These preliminary results based on a small sample of only 9
galaxies will be put on a much more solid ground, both from the
statistical and physical point of view, by the forthcoming analysis of
a larger sample of galaxies for which multiwavelength observations,
from UV to radio, are available.

\end{document}